\documentclass[twocolumn,aps,amsmath,amssymb,prl]{revtex4}
\usepackage{epsfig,bm,dcolumn}
\usepackage{graphicx}
\newcommand {\beq} {\begin{equation}}
\newcommand {\eeq} {\end{equation}}
\newcommand {\ca} {\ensuremath{c^{\dagger}}}
\newcommand {\up}  {\ensuremath{\uparrow}}
\newcommand {\dn}  {\ensuremath{\downarrow}}
\newcommand {\bqa} {\begin{eqnarray}}
\newcommand {\eqa} {\end{eqnarray}}

\newcommand {\tr} {\ensuremath{t_{{\bf rr'}}}}
\newcommand {\rr} {\ensuremath{{\bf r}}}
\newcommand {\rp} {\ensuremath{{\bf r'}}}
\newcommand {\kk} {\ensuremath{{\bf k}}}
\newcommand {\kp} {\ensuremath{{\bf k'}}}

\begin{document}
\topmargin 0.1cm
\title{Can one determine the underlying Fermi surface in the superconducting state of strongly correlated systems?}
\author{Rajdeep Sensarma}
\author{Mohit Randeria} 
\author{Nandini Trivedi}
\affiliation{Department of Physics, The Ohio State University, Columbus, Ohio 43210}
\begin{abstract}
The question of determining the underlying Fermi surface (FS) that is gapped by superconductivity (SC) is of central
importance in strongly correlated systems, particularly in view of angle-resolved photoemission experiments.
Here we explore various definitions of the FS in the superconducting state
using the zero-energy Green's function,  the excitation spectrum and the momentum distribution.
We examine (a) d-wave SC in high Tc cuprates, and 
(b) the s-wave superfluid in the BCS-BEC crossover. In each case we show that the various definitions 
agree, to a large extent, but all of them violate the Luttinger count and do not enclose 
the total electron density. We discuss the important role of
chemical potential renormalization and incoherent spectral weight in this violation.
\end{abstract}
\maketitle 


The Fermi surface (FS), the locus of gapless electronic excitations in ${\bf k}$-space,
is one of the central concepts in the theory of Fermi systems.
In a Landau Fermi liquid at T$=$0, Luttinger \cite{Luttinger} defined the FS in terms of 
the single-particle Green's function $G^{-1}(\kk,0) = 0$ and showed that it encloses the same volume as in 
the non-interacting system, equal to the fermion density $n$.
In many Fermi systems of interest the ground state has a broken symmetry. Here we study states with 
superconducting (SC) long-range order, where there is no surface of gapless excitations, and
ask the question: Is there any way to define at T$=$0 the ``underlying Fermi surface'' that 
got gapped out by superconductivity?

>From a theoretical point of view, this question is of relevance to all superconductors
irrespective of pairing symmetry or mechanism. The answers turn out to be of particular interest
for strongly correlated superconductors, where the surpring effects that we find are large
enough to be measured experimentally. Angle-resolved photoemission spectroscopy (ARPES) \cite{Arpes}
has emerged as one of the most powerful probes of complex materials and has been extensively
used to determine the FS in strongly correlated systems, often
from data in the SC state \cite{SCstateFS,Fujimori}. One of our goals is to understand exactly what a T$=$0 measurement can tell us
about the FS. This is especially important in the cuprates where the
the normal state must necessarily be studied at high temperatures and does not show sharp electronic excitations, 
expected in Fermi liquids, in contrast to the SC state which does show sharp Bogoliubov quasiparticles.
Our results are also of interest for a completely different class of systems: strongly interacting 
Fermi atoms \cite{ColdAtoms} in the BCS-BEC crossover \cite{Leggett,MRreview}. Here too the
question of an underlying Fermi surface is of direct experimental relevance \cite{Jin}.

In this paper, we first show that Luttinger's original argument \cite{Luttinger} 
cannot be generalized to the SC state, and this violation
is related to broken gauge invariance \cite{Dzyaloshinski}.
We then explore various criteria for defining the ``underlying Fermi surface'' in the T$=$0 SC state, 
using properties of the single-particle Green's function $G(\kk,\omega)$ \cite{Nambu}
directly related to experimentally measurable quantities. We present results for the two
systems described above:
(a) the d-wave SC state of the high Tc cuprates which is dominated by strong Coulomb correlations, 
and (b) the s-wave superfluid state in the BCS-BEC crossover regime of atomic Fermi gases with 
strong attractive interactions.
We will show that the various definitions lead to FS contours which are not identical, 
but nevertheless agree with each other to a remarkable degree.
All of them violate the Luttinger sum rule (area enclosed equal to fermion density) and we 
obtain a detailed understanding of this violation: its magnitude is related to the
SC gap function and its sign to the topology of the FS. 

{\bf Fermi surface criteria:}
It is perhaps not appreciated that the question of the ``underlying FS'' 
in the SC state is non-trivial, because in BCS theory the answer appears to be simple. 
In the BCS state one can look at \cite{Nambu} 
$G^{\rm BCS}(\kk,z) = (z + \xi_\kk)/(z^2 - E_\kk^2)$ and ask where $G^{\rm BCS}(\kk,0) = 0$.
The resulting surface coincides with $\xi_\kk = 0$, the \emph{normal state} FS on which the 
pairing instability takes place. Thus it is tempting to use $G(\kk,0) = 0$ in a more general setting to define the 
SC state FS. This is analogous to Luttinger's definition except
$G(\kk,0)$ changes sign through a zero in the SC, instead of a pole in the normal case.
However, it is important to note \cite{Dzyaloshinski} that there is no analog of Luttinger's theorem for SCs.
One can write the Luttinger-Ward functional in terms of the Nambu Green's function matrix ${\hat{\bf G}}$,
and try to generalize Luttinger's proof \cite{Luttinger}. However ${\rm Tr}{\hat{\bf G}}$ only constrains the 
\emph{difference} $(n_\up - n_\dn)$, which is trivially zero in our case \cite{Sachdev}, and not the sum \cite{tau3}.
This is related to the fact that spin $S_z^{\rm total}$ is conserved in the SC state but number is not.  
Thus one \emph{cannot} show in general that the surface \cite{Nambu} $G(\kk,0) = 0$ in the SC state encloses $n$ fermions. 
We will come back later to why the Luttinger count nevertheless \emph{seems} to work in BCS theory. 

We next turn to various alternative definitions of the FS.
(i) ARPES measures \cite{Randeria} the one-particle spectral function 
$A({\bf k},\omega) = - {\rm Im}G({\bf k},\omega+ i0^+)/\pi$ and thus one way to define the
``underlying FS'' is to look at $A(\kk,\omega=0)$ to map out the locus of maximum ARPES intensity.
We also describe below a closely related minimum gap locus, also motivated by ARPES experiments 
\cite{Arpes,SCstateFS}. (ii) In the cold-atom experiments \cite{Jin}, it is possible to measure the 
momentum distribution, and therefore we also discuss the (somewhat ad-hoc but well defined) criterion $n(\kk) = 1/2$ 
to define a surface that separates states of high and low occupation probabilities.
(iii) We show below that the quasiparticle excitation spectrum, even in the strongly correlated SC state, 
is given by $E_{\kk} = (\xi_{\kk}^2+|\Delta_{\kk}|^2)^{1/2}$ where $\xi_{\kk}$ is the \emph{renormalized} dispersion and
$\Delta_{\kk}$ the gap function. We then look at the contour defined by $\xi_\kk = 0$ to define the ``FS''.
In addition to comparing the contours obtained by various definitions, we also discuss the extent to which
these results differ from $G(\kk,0) = 0$.

\begin{figure}[h!]
\centering
\includegraphics[scale=0.30,angle=270]{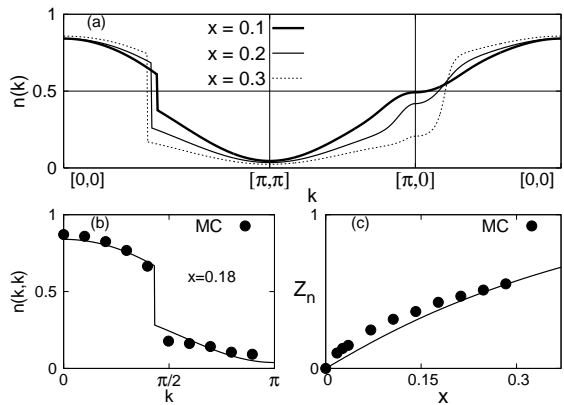}
\caption{(a): Momentum distribution $n({\kk})$ along 
$(0,0) \rightarrow (\pi,\pi) \rightarrow (\pi,0) \rightarrow (0,0)$
for three different doping levels. Comparison of renormalized mean field theory
with variational Monte Carlo (MC) results of ref.~\cite{Paramekanti}: 
(b) $n({\kk})$ and (c) nodal $Z({\kk})$ as function of $x$.
}
\label{nkcomp}
\end{figure}

{\bf (a) High Tc Superconductors:} We describe the strongly correlated d-wave superconducting
ground state and low-lying excitations using a variational approach \cite{RVB,Paramekanti} 
to the large $U$ Hubbard model
${\cal H}=-\sum_{\rr \rp}\tr \ca_{\rr \sigma}c_{\rp \sigma}+U\sum_{\rr} n_{\rr \up}n_{\rr \dn}$ on a 2D square lattice.
We choose \cite{Parameters} the bare dispersion 
$\epsilon_{\kk}= - t\gamma_{\kk} + t^\prime \lambda_{\kk}$ with $\gamma_{\kk}=2(\cos k_x + \cos k_y)$ and 
$\lambda_{\kk}=4\cos k_x \cos k_y$. We work at an electron density $n = 1 -x$ with hole doping $x \ll 1$. 
Our variational ground state is 
$|\psi_0 \rangle = \exp(-i{\cal S}){\cal P}|BCS \rangle$, where
$|BCS \rangle$ is the BCS wavefunction with $d_{x^2-y^2}$ pairing,
the projection operator ${\cal P}=\prod_{\rr}(1-n_{\rr\up}n_{\rr\dn})$ 
eliminates all double-occupancy and finite $t/U$ corrections
are built in through $\exp(-i{\cal S})$ \cite{Paramekanti}.
Here we present the results of a renormalized mean field theory (RMFT)
using the Gutzwiller approximation \cite{GutzwillerApprox}
which are in excellent agreement (see Fig.~\ref{nkcomp}(b,c)) 
with those obtained using the variational Monte Carlo (MC) \cite{Paramekanti} which treats
projection exactly. The RMFT approach has advantages over MC for our present
investigation since we get much better ${\bf k}$-resolution and we
can study spectral functions.

In the RMFT we minimize $\langle {\cal H} - \mu N \rangle $ to
obtain self-consistency equations for the gap function 
$\Delta_{\kk}= \Delta(\cos k_x-\cos k_y)/2$ and Fock 
shift $\chi_{\kk}$ \cite{HartreeFock}. These in turn determine the BCS factors
$v_\kk^2 = 1 - u_\kk^2 = (1 - \xi_\kk/E_\kk)/2$.
The renormalized dispersion
$
\xi_{\kk}= g_t\epsilon_{\kk}- \chi_{\kk} - \mu
$
incorporates bandwidth suppression by the Gutzwiller
factor $g_t=2x/(1+x)$, the Fock shift $\chi_{\kk}$, and
the (Hartree shifted) chemical potential $\mu$. 
$
E_{\kk} = (\xi_{\kk}^2+|\Delta_{\kk}|^2)^{1/2}
$
is the excitation energy \cite{GutzwillerApprox} of 
the projected Bogoliubov quasiparticle (QP) state
$|\kk\up\rangle=\exp(-i{\cal S}){\cal P}\gamma^{\dagger}_{\kk\up}|BCS \rangle$,
where $\gamma^{\dagger}_{\kk\up}=(u_{\kk}\ca_{\kk\up}-v_{\kk}c_{-\kk\dn})$. 

The spectral function is of the form \cite{Rajdeep} 
$A(\kk,\omega) = A^{\rm coh}(\kk,\omega) + A^{\rm inc}(\kk,\omega)$.
The coherent part
$A^{\rm coh}(\kk,\omega)=Z(\kk)\left[u_{\kk}^2\delta(\omega-E_{\kk}) + v_{\kk}^2\delta(\omega+E_{\kk})\right]$
where the quasiparticle residue
$Z(\kk)=g_t - 2g_{st}\sum_{\kp}\epsilon_{\kp}v_{\kp}^2/U - 2g_t\epsilon_{\kk}\sum_{\kp}v_{\kp}^2/U$
with $g_t = 2x/(1+x)$ and $g_{st} = 4x/(1+x)^2$.
The coherent weight $Z$ decreases monotonically with underdoping, vanishing as $x \to 0$,
as seen in Fig.~\ref{nkcomp}(c), while the \emph{incoherent} spectral weight 
$A^{\rm inc}$ increases with decreasing $x$ as required by rigorous sum rules \cite{Rajdeep}.

\begin{figure}[h!]
\centering
\includegraphics[scale=0.65,angle=270]{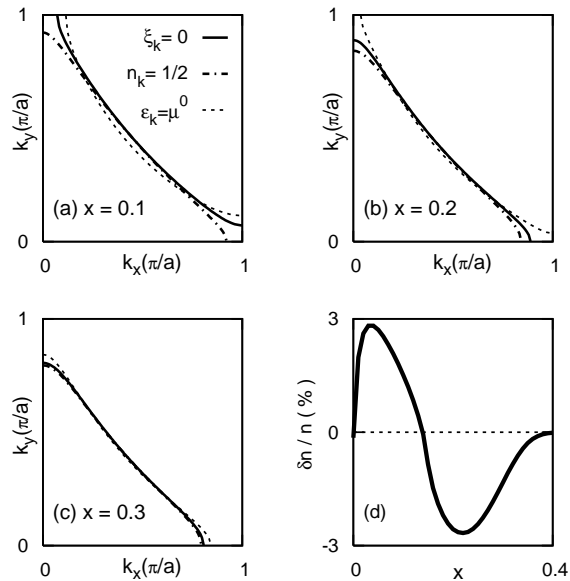}
\caption{(a,b,c): Various ``Fermi surface Contours'' as a function of hole doping $x$.
The contours plotted are (1) $\xi_\kk = 0$, (2) $n(\kk) = 1/2$, and  (3) the 
non-interacting Fermi surface $\epsilon_\kk = \mu^0$. Note that the ``minimum gap locus'' 
(see text) is indistinguishable from $\xi_\kk = 0$.
(d): The fractional difference $\delta n/n$ between the area enclosed by 
$\xi_\kk = 0$ and $n$ plotted as a function of hole doping $x$.
}
\label{fscontours}
\end{figure}

{\bf Renormalized dispersion:} The form of the excitation gap $E_{\kk} = (\xi_{\kk}^2+|\Delta_{\kk}|^2)^{1/2}$
suggests that we identify $\xi_\kk = 0$ as the ``underlying FS''. We must emphasize, that despite the BCS-like
form of $E_{\kk}$, \emph{the theory neither assumes nor implies the existence of sharp quasiparticles
in the normal state with a dispersion} $\xi_\kk$. In Fig.~\ref{fscontours} we plot the $\xi_\kk = 0$ contours
for various $x$ along with others which will be discussed below. 
We see that the $\xi_\kk = 0$ contours are hole-like -- closed around $(\pi,\pi)$ -- for small $x$, but
electron-like -- closed around $(0,0)$ -- for $x > 0.16$. 
The precise $x$ at which the ``FS'' topology changes is a sensitive function of 
the bare parameter \cite{Parameters} $t'/t$.

\begin{figure}[h!]
\includegraphics[scale=0.24]{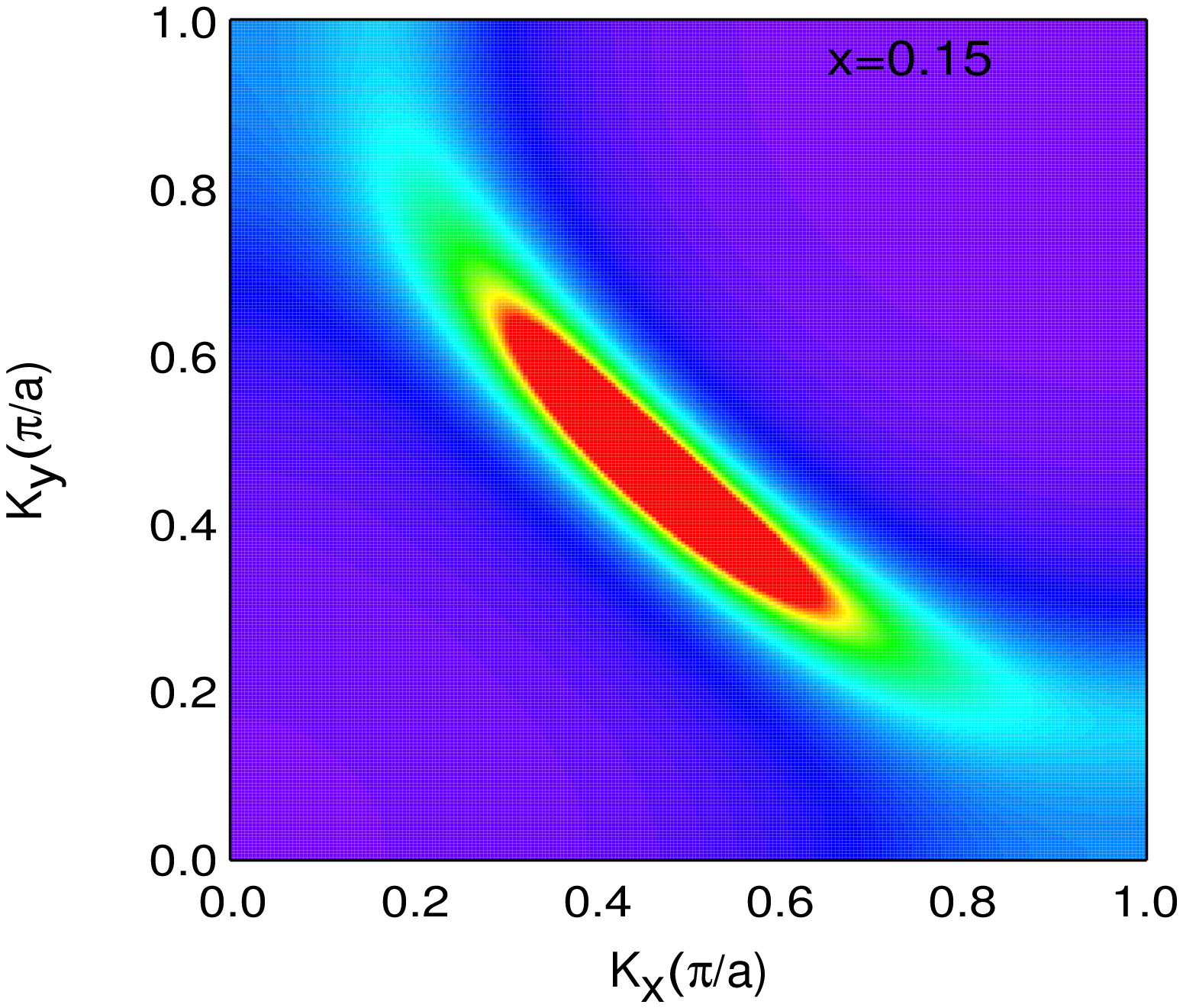}
\includegraphics[scale=0.24]{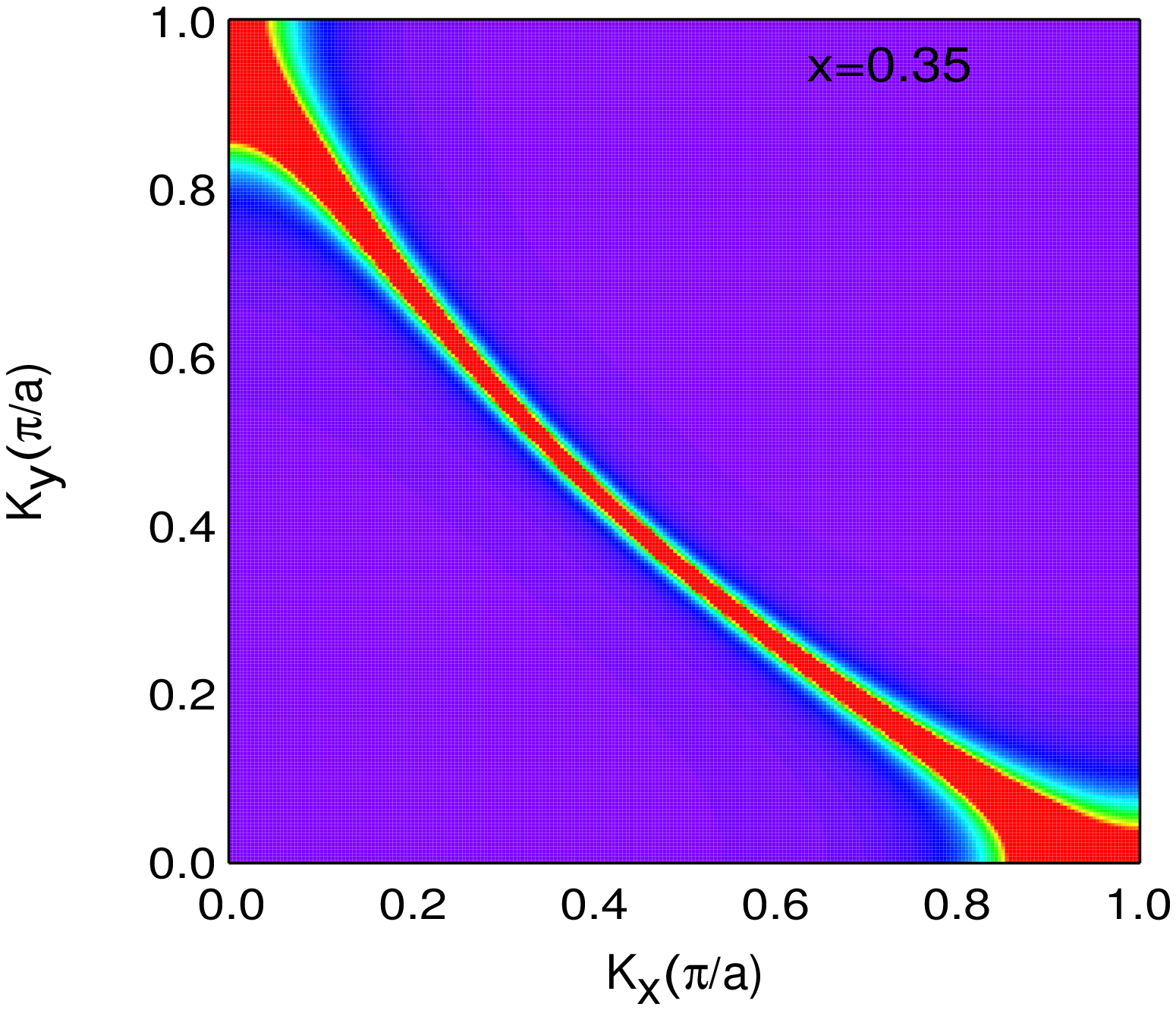}
\caption{(Color online) ARPES intensity maps at zero energy $A(\kk,0)$ for $x=0.15$ (left) and $x=0.35$ (right).
}
\label{fs-intensity}
\end{figure}

{\bf Minimum gap locus:} We plot in Fig.~\ref{fs-intensity}
$A(\kk,0)$ which is the zero energy ARPES intensity. 
We can neglect the incoherent weight at $\omega = 0$ and write
$A(\kk,0) \simeq Z(\kk)\Gamma/(E_\kk^2 + \Gamma^2)$,
where the $\delta$-function is broadened with a small $\Gamma$.   
We now follow a procedure developed
in analyzing ARPES experiments \cite{Arpes}. Various cuts through 
${\kk}$-space are taken perpendicular to $\xi_{\kk}= 0$. On each of these cuts
we determine the location of the maximum $A(\kk,0)$, which is also
the same as minimum gap $E_{\kk}$ (ignoring
the negligible $\kk$-dependence of $Z(\kk)$ on this locus).
The locus of $\min E_{\kk}$ defines the ``minimum gap locus''.
We see from Fig.~\ref{fscontours} that the curve $\xi_{\kk}= 0$ and the 
``minimum gap locus'', although not identical, are very similar at every doping. 
In fact this difference is not visible in Fig.~\ref{fscontours} as it 
is less than the width of the lines used. 

{\bf Momentum distribution:} 
We calculate $n({\kk})= \langle c^\dag_{\kk\sigma}c_{\kk\sigma}\rangle$
and find the result $n({\kk}) = Z(\kk) v_{\kk}^2+ \tilde{n}(\kk)$,
where first term comes from coherent quasiparticles and the second term
$\tilde{n}(\kk)=(1-x)^2/2(1+x) + {\cal O}(t/U)$ is the incoherent contribution 
\cite{nk-footnote}. The evolution of $n_{\kk}$ with doping $x$ is shown in 
Fig.~\ref{nkcomp}(a). We choose the contour $n_{\kk} = 1/2$ \cite{One-half} to
(somewhat arbitrarily) separate states of high and low occupation and its
variation with $x$ is plotted in Fig.~\ref{fscontours}. We see that
unlike $\xi_\kk = 0$, the $n(\kk) = 1/2$ contour does not exhibit a change
in topology and is `electron-like' down to very low doping.
This qualitative difference arises because $\xi_\kk = 0$ and the ``minimum gap locus''
depend only on the coherent part of the spectral function, while $n(\kk)$ is an energy-integrated quantity
that includes incoherent spectral weight \cite{pole-zero}. 

{\bf Luttinger Count:} 
We see from Fig.~\ref{fscontours} that the various ``FS'' contours 
enclose areas different from $n$, which is the area enclosed by  
the \emph{non-interacting} FS $\epsilon_\kk = \mu^0$.
In Fig.~\ref{fscontours}(d)
we plot the difference between the area enclosed by $\xi_{\kk}= 0$ 
and $n$ as a function of hole doping $x$. We see that, in general, this difference is non-zero 
when the system exhibits SC long range order ($0 < x < 0.4$) \cite{Mott}.
For $x < 0.16$, the $\xi_{\kk}= 0$ contour is hole-like and we find an
area enclosed greater than $n$, while for $0.16 < x < 0.4$, this contour is electron-like and 
the enclosed area is less than $n$. The $x > 0.4$ ground state is a normal 
Fermi liquid and the Luttinger sum rule is valid \cite{fermi-liq-limit}. 

A simple way to understand the variations seen in Fig.~\ref{fscontours}(d), which include both Gutzwiller and 
Hartree-Fock renormalizations, is not immediately obvious. 
However, the BCS-BEC crossover analysis below will
give us clear insight into both the (small) magnitude of the violation observed here and the relation
of its sign to the ``FS'' topology.

{\bf Zeros of G:}
Next we compare the FS contours obtained above with the surface ${\textsl G}(\kk,0) = 0$,
though we note that the latter is not of direct experimental relevance.
>From the form of $A(\kk,\omega)$ obtained above, we get
\beq
G(\kk,0) = - Z_\kk \xi_\kk / E_\kk^2 
- P\int_{-\infty}^{+\infty} d\omega A^{\rm inc}(\kk,\omega) / \omega.
\label{green-fn}
\eeq 
It is clear that $\xi_\kk = 0$  corresponds only to
the first term ${\textsl G}^{\rm coh}(\kk,0) = 0$, and \emph{not} \cite{pole-zero} to a zero of 
the full ${\textsl G}$. From the sum-rule constraints \cite{Rajdeep}
on $A^{\rm inc}$ one can show that the integral above is necessarily negative.
It then follows that the zeros of ${\textsl G}$ correspond to 
$\xi_\kk > 0$. This implies that for small $x$, where $\xi_\kk = 0$ is
hole-like, ${\textsl G} = 0$ gives an even larger violation of Luttinger count. 

{\bf (b) BCS-BEC Crossover}: The evolution of a Fermi gas from a 
BCS paired superfluid to a BEC of composite bosons has now been realized in the laboratory \cite{ColdAtoms}
using a Feshbach resonance to tune the s-wave scattering length $a_s$. The dimensionless coupling
$g = 1/k_f a_s$ can be varied from large negative (BCS limit) to large positive (BEC limit) values, with
unitarity ($g = 0$) being the most strongly interacting point in the crossover. 
We use the T$=$0 crossover theory \cite{Leggett,MRreview} to gain further insight into
the question of the ``underlying Fermi surface''.

The structure of the T$=$0 Green's function \cite{Nambu} in this case
is exactly the same as eq.~(\ref{green-fn}), with an excitation spectrum 
$E_{\kk} = (\xi_{\kk}^2+ \Delta^2)^{1/2}$. In the well-controlled limit of
large dimensionality, treated within dynamical mean field theory \cite{Garg}, $Z$ is close to unity and the
incoherent spectral weight is small even at unitarity. Thus, to make our point in the simplest possible manner,
we work with Leggett's variational ansatz \cite{Leggett,Engelbrecht}.
Even at this level, where incoherent weight vanishes, the implications of the various FS definitions are very interesting. 

It is then easy to show analytically \cite{crossover-footnote} that \emph{all} the definitions investigated above, 
yield the \emph{same surface in} $\kk$-space \emph{for all values
of the coupling} $g=1/k_f a_s$. 
This is given by the bare dispersion $\hbar^2 k^2/2m = \mu(g)$, where 
the chemical potential $\mu$ \emph{strongly renormalized} from its non-interacting value. 
Even in the weak-coupling BCS limit, $\mu$ is less than the non-interacting $\epsilon_f$ by an exponentially
small amount of order $\Delta^2/\epsilon_f$ and the Luttinger count is violated. This violation becomes increasingly
severe with increasing $g$: as the gap increases, $n(\kk)$ broadens and $\mu$ decreases; see ref.~\cite{Engelbrecht}.
Eventually on the BEC side of unitarity ($g \sim 1$), $\mu$ goes negative and the surface
$\xi_\kk = 0$ shrinks to $\kk=0$, beyond which one enters the Bose regime. 
To summarize: the ``underlying FS'' does \emph{not} enclose the total number density $n$, 
its volume decreases monotonically with $g$ and for $g$ greater than a critical value 
it is zero! 

This analytical solution is modified quantitatively by correlation effects beyond the Leggett theory, 
but qualitative effects like the decrease in $\mu$ and broadening
of $n(\kk)$ with increasing $g$ persist, as also seen in both numerical \cite{Qmc} and 
experimental studies of $n(\kk)$ \cite{Jin}. 

We now see how the renormalization of the chemical potential in the presence of a SC condensate
directly leads to a violation of the Luttinger count. For not too large attraction, the 
violation has a relative size $(\Delta/\epsilon_f)^2$, and a negative sign for a particle-like FS,
i.e., the underlying FS encloses a smaller area than the non-interacting FS.
To see how the sign changes for a hole-like FS,we look at the BCS-BEC crossover in 
a lattice model such as the attractive Hubbard Hamiltonian \cite{Lotfi}. It is straightforward to show,
using a particle-hole transformation, that sign of the $\mu$-renormalization reverses going from a
particle-like to a hole-like FS. Thus we find that for a hole-like FS, the underlying FS in the SC state
encloses a larger area than the bare FS. These are exactly the effects seen 
in the strongly correlated d-wave SC in Fig.~\ref{fscontours}(d).

{\bf Conclusions:}  We have analyzed various criteria for the ``underlying FS''
in the T$=$0 SC state. We have shown that a ``FS'' deduced from a SC state measurement
necessarily violates the Luttinger sum rule and does not, in general, enclose $n$ fermions.
We have gained detailed insights into the magnitude and sign of the violation. Our results are 
of most interest for the high Tc cuprates, where they can be tested in ARPES experiments, provided one
can independently determine the electron density. The existing ARPES data \cite{Fujimori,Oxychloride} on LaSrCuO
show a small violation of the Luttinger count with a sign change, consistent with our results.
We should emphasize that our theoretical considerations makes no statement about the finite temperature 
non-Fermi liquid normal states.

{\bf Acknowledgments}
We acknowledge useful discussions with J.C. Campuzano, A. Fujimori and S. Sachdev.
We thank P.W. Anderson for sending us a copy of ref.~\cite{Gros} while we were preparing this 
manuscript. The results of ref.~\cite{Gros} are very similar to those in part (a) of our paper.


%
%
%

\begin{thebibliography}{99}

\bibitem{Luttinger}
J. Luttinger, Phys. Rev. {\bf 119}, 1153 (1960); A. A. Abrikosov, L. P. Gorkov and I. Dzyaloshinski,
{\it Methods of Quantum Field Theory in Statistical Physics}, (Dover, NY, 1963); Sec.~19.4.

\bibitem{Arpes} A. Damascelli, Z. Hussain and Z. X. Shen, Rev. Mod. Phys. {\bf 75}, 473 (2003);
J. C. Campuzano, M. R. Norman and M. Randeria, in `Physics of Superconductors', Vol. II, K. Bennemann and
J. Ketterson eds. (Springer, Berlin, 2004).

\bibitem{SCstateFS}
J. C. Campuzano, {\it et al.}, Phys.~Rev.~B {\bf 53} R14737, (1996);
H. Fretwell, {\it et al.}, Phys. Rev. Lett. {\bf 84}, 4449 (2000).

\bibitem{Fujimori}
T. Yoshida, {\it et al.}, cond-mat/0510608.

\bibitem{ColdAtoms} M. Greiner, C. A. Regal, and  D. S. Jin, Nature {\bf 426}, 537 (2003);
M. W. Zwierlein, {\it et al.}, Phys. Rev. Lett. {\bf 91}, 250401 (2003).
 
\bibitem{Leggett}
A. J. Leggett in {\it Modern Trends in the Theory of Condensed Matter},
A. Pekalski and R. Przystawa, eds. (Springer, Berlin, 1980).

\bibitem{MRreview} M. Randeria in {\it Bose-Einstein Condensation}, A. Griffin, D. Snoke and
S. Stringari, eds. (Cambridge 1995).

\bibitem{Jin} C~A. Regal, {\it et al.}
Phys. Rev. Lett. 95, 250404 (2005).

\bibitem{Dzyaloshinski}
I. Dzyaloshinskii, Phys. Rev. B. {\bf 68}, 085113 (2003).

\bibitem{Nambu}
Here $G$ is the ${\hat{\bf G}}_{11}$ element of the Nambu matrix ${\hat{\bf G}}$.

\bibitem{Randeria} 
M. Randeria, {\it et al.}, Phys Rev. Lett. {\bf 74}, 4951 (1995).


\bibitem{Sachdev} It can be important for \emph{unequal} spin populations; see S. Sachdev and K. Yang, cond-mat/0602032.

\bibitem{tau3}
$(n_\up + n_\dn)$ is related to ${\rm Tr}\{\hat{\bf G}\hat{\tau_3}\}$, 
for which the Luttinger argument \cite{Luttinger} fails. 

\bibitem{RVB} P.W. Anderson, Science {\bf 235}, 1196 (1987). 

\bibitem{Paramekanti} A. Paramekanti, M. Randeria and N. Trivedi, Phys. Rev. Lett, {\bf 87}, 217002 (2001)
and  Phys. Rev. B. {\bf 70}, 054504 (2004).

\bibitem{Parameters}
For our numerical results we choose typical values for the cuprates: $t=300$ meV, $t'= t/4$ and $U/t = 12$ 
corresponding to $J = 4t^2/U = 100$ meV. 

\bibitem{GutzwillerApprox} 
F.C. Zhang, {\it et al.},
Sup. Sci. Tech. {\bf 1}, 36 (1988);
P.W. Anderson, {\it et al.},
J. Phys. Cond. Mat. {\bf 16}, R755 (2004).

\bibitem{HartreeFock}
We find that
$\chi_{\kk}=\xi_1\gamma_{\kk}+ \xi_2\lambda_{\kk} +\xi_3\gamma^2_{\kk}+\xi_4\lambda^2_{\kk}+
\xi_5\gamma_{\kk}\lambda_{\kk}$, where $\gamma_{\kk}$ and $\lambda_{\kk}$ were used to
define $\epsilon_\kk$. 


\bibitem{Rajdeep} M. Randeria, {\it et al.},
Phys. Rev. Lett. {\bf 95}, 137001 (2005).

\bibitem{nk-footnote}
The full expression will be published separately; R. Sensarma, M. Randeria and N. Trivedi (unpublished).

\bibitem{One-half}
On the zone diagonal $n_{\kk}$ has a jump discontinuity of $Z(\kk_n)$ at the nodal $\kk_n$; see
Fig.~\ref{nkcomp}. The FS crossing on the diagonal is defined by $\kk_n$, 
rather than $n(\kk) = 1/2$.

\bibitem{Mott}
The Mott insulator \cite{Paramekanti} at $x=0$ has $Z \equiv 0$ and
is outside the scope of discussion here. 

\bibitem{pole-zero}
Note that for a normal Fermi liquid, it follows from the analog of eq.~(\ref{green-fn}) that incoherent spectral 
weight has no effect on the Luttinger FS $G^{-1}(\kk,0) = 0$.

\bibitem{fermi-liq-limit} In the $\Delta = 0$ Fermi-liquid limit, one can show analytically from
that all the contours are identical.

\bibitem{Garg}
A. Garg, H. R. Krishnamurthy and M. Randeria, Phys. Rev. B {\bf 72}, 024517 (2005).

\bibitem{Engelbrecht}
J. R. Engelbrecht, M. Randeria and C. Sa de Melo, Phys. Rev. B {\bf 55}, 15153 (1997).

\bibitem{crossover-footnote}
From \cite{Engelbrecht} we see
$G(\kk,0) = - \xi_k/E_\kk^2 = 0$ and $n(\kk) = [1 - \xi_k/E_\kk]/2 = 1/2$ are the same
as $\xi_\kk =0$, as is the minimum gap locus deduced from $E_\kk$. 
Inclusion of the effects of $A^{\rm inc}$ in (\ref{green-fn}) will lead to differences
between the various surfaces though, as discussed in case (a).

\bibitem{Qmc} G.~E. Astrakharchik, {\it et al.}
Phys. Rev. Lett. {\bf 95}, 230405 (2005).

\bibitem{Lotfi}
L. Belkhir and M. Randeria, Phys.~Rev.~B {\bf 45}, 5087 (1992).

\bibitem{Oxychloride} 
Our theory does not apply to the charge ordered SC state
of NaCaCuOCl (H. Hanaguri {\it et al.}, Nature {\bf 430} 1001 (2004))
where ARPES sees a larger violation \cite{Fujimori}.

\bibitem{Gros} 
C. Gros, 
B. Edegger, V. N. Muthukumar and P. W. Anderson, 
cond-mat/0606750.

\end{thebibliography}
\end{document}